\documentclass[onecolumn,aps,prd,amsmath,amssymb,floatfix,amsmath,superscriptaddress]{revtex4-1}

\usepackage{graphicx}% Include figure files
\usepackage{dsfont}
\usepackage{dcolumn}% Align table columns on decimal point
\usepackage{bm}% bold math
\usepackage{textcomp}
\usepackage{wasysym}
\usepackage{stmaryrd}
\usepackage{slashed}
\usepackage{epsfig}
\usepackage{xcolor}
\usepackage{hyperref}

\def \ee{\end{equation}}
\def \be{\begin{equation}}
\def \eea{\end{eqnarray}}
\def \bea{\begin{eqnarray}}

\begin{document}

\title{Earthquake Quantization
%\\
%\vspace{0.3cm}
%Non-relativistic Point Particle
%Path Integal or Curvature Integral 
%the only thing we have to do
}

\author{Benjamin Koch}
 \email{benjamin.koch@tuwien.ac.at}
% \affiliation{Institute for Theoretical Physics, TU Wien, Wiedner Hauptstr. 8, A-1040 Vienna, Austria}
\affiliation{Institut f\"ur Theoretische Physik and Atominstitut,
 Technische Universit\"at Wien,
 Wiedner Hauptstrasse 8--10,
 A-1040 Vienna, Austria}
\affiliation{Facultad de F\'isica, Pontificia Universidad Cat\'olica de Chile, Vicu\~{n}a Mackenna 4860, Santiago, Chile}
 
\author{Enrique Mu\~{n}oz}
\email{munozt@fis.puc.cl}
\affiliation{Facultad de F\'isica, Pontificia Universidad Cat\'olica de Chile, Vicu\~{n}a Mackenna 4860, Santiago, Chile}
% \altaffiliation[Also at ]{Physics Department, XYZ University.}%Lines break automatically or can be forced with \\
%\altaffiliation[Also at]{
%FIAS--Frankfurt Institute for Advanced Studies, \\
%Max von Laue-Str. 1,D--60438 Frankfurt am Main, Germany\\
%}
\date{\today}

\begin{abstract}
%In this homage to Einstein's 144th birthday w
We propose a novel perspective on quantization, where the paths of a path-integral are not random, but rather
solutions of a geodesic equation
in a random background. We show that
this change of perspective can be made mathematically equivalent to the usual formulations of non-relativistic quantum mechanics. 
To conclude, we comment
 on conceptual issues, such as 
 quantum gravity coupled to matter and 
 the quantum equivalence principle.
\end{abstract}

\maketitle

\tableofcontents

%%%%%%%%%%%%%%%%%%%%%%%%%%%%%%%55
\section{Prologue}
%%%%%%%%%%%%%%%%%%%%%%%%%%%%%%%%%%%%%%%%%%%%%%%%%%
%https://youtu.be/Q2pw-KdQtso

%Usually, the quantum mechanical path integral~\cite{Feynman:1950ir} is constructed from a sum over all random paths which connect two points in space and time.
%Those paths follow no law of motion, they are postulated and then weighted by their action and integrated over with a suitable measure.

%Let's exemplify our idea with a pictorial analogy:
Given the occasion, it seems appropriate to start with an analogy in terms of a thought experiment:\\
Imagine that you are sitting on a balloon and observing the motion of few people walking on a square. Surprisingly,
you do not see a smooth flow but rather
a random zigzag.
When trying to make sense of this crazy motion you might conclude that
the people are
\begin{itemize}
    \item[a)]  actually drunk and are thus moving strangely on their own cause;
    %This interpretation might not satisfy you, since it would imply that in this strange town nobody is sober.
    \item[b)] 
      sober, but they have a hard time trying to move steadily since they are suffering from a massive earthquake, which you in your safe balloon, can not perceive directly.
\end{itemize}
The first alternative  a) of this analogy corresponds to intrinsic random motion of the path integral (PI) quantization~\cite{Feynman:1950ir}. 
The second alternative b) corresponds to random motion caused by a random background.

The idea of this short comment is that the paths of the PI could actually not be random 
by themselves, instead they are the
result of a law of motion. Different paths are imposed by different causes
just like the poor people in the above analogy.
We will explore whether it is possible to cast the sum over random paths in the form of a sum over 
``classical'' paths driven by random causes.
In classical physics, deflections from straight paths are associated to forces. 
There are four known fundamental forces in nature: the electromagnetic-, weak-, strong-, and gravitational force.
We do not want to introduce an entirely new force to implement the above idea, thus we have to resort to one of these four known forces. When making this choice, we recall the fact that the laws of quantum mechanics are universal in the sense that all fields are quantized, regardless of their charge or mass.
In terms of the path integral formulation, this universality means that one has to integrate over all field configurations irrespective of their charge or their mass. This observation singles out gravity as the only interaction which is universal in the same sense: ``Geodesics of test particles are independent of the particles charge and mass''. This is the reason why in the following discussion we resort to gravity
and not to other forces.
We will show that
non-relativistic quantum mechanics can be expressed as weighted sum over ``quantum-geodesics'' resulting from fluctuations of the background metric.
Because of the above analogy we call this approach earthquake quantization (EQQ)
\cite{EQQ:2023}.
%\footnote{Note that this name is not meant as trivialization of the devastating effects an earthquake can provoke in inhabited regions. Instead, we encourage everyone to donate to those who are helping in the aftermath of such an event.}.

%%%%%%%%%%%%%%%%%%%%%%%%%%%%%%%55
\section{Introduction}

%%%%%%%%%%%%%%%%%%%%%%%%%%%
\subsection{What is a trajectory?}

\begin{figure}[hbt]
   \centering
\includegraphics[width=12cm]{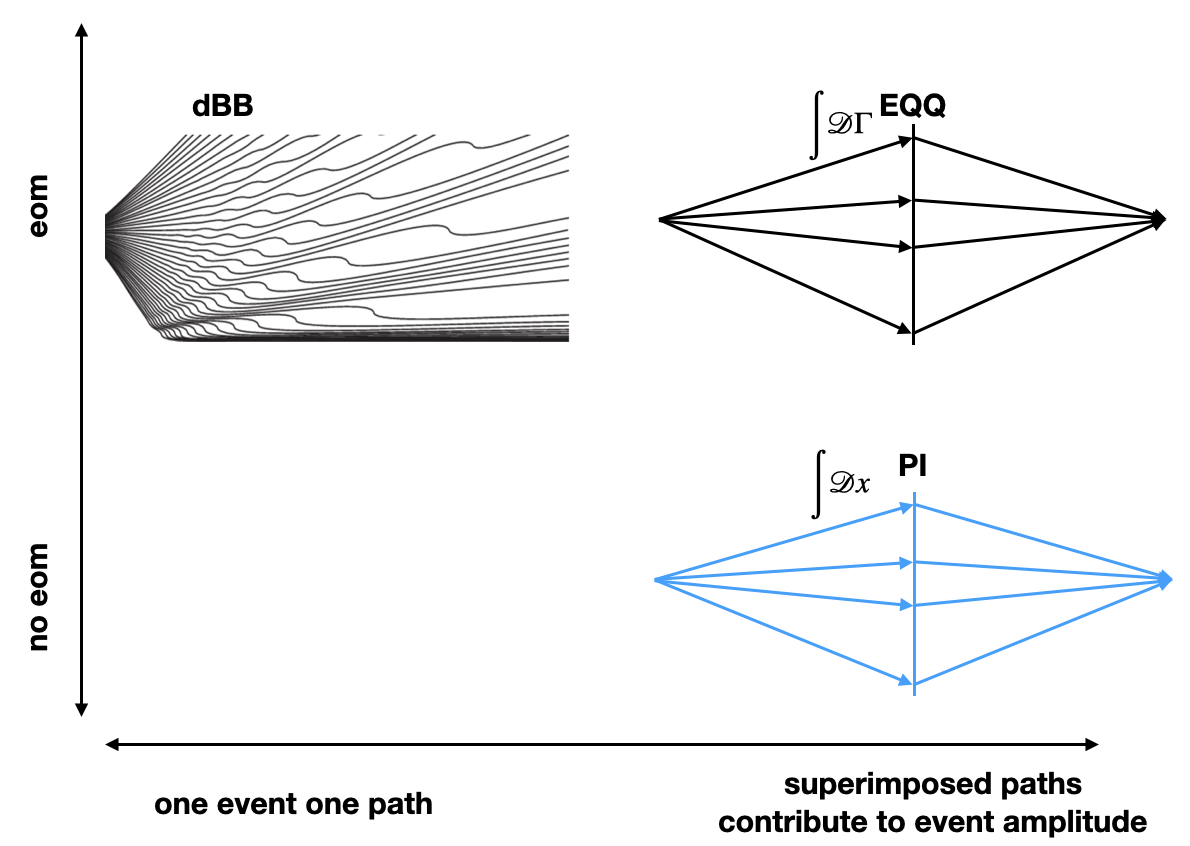}
  \caption{
  Conceptual map, showing the similarities and differences between dBB, PI, and EQQ. Blue paths are ``drunken'' (no equation of motion (eom)), black paths are ``sober'' (with an eom).
  \label{fig_BMPIEQ}}
\end{figure}

A classical trajectory (TJ) of a point particle associates to every given time a position. Throughout the history of physics, this simple concept did suffer several 
serious contraventions such as:
Particles are not point-like, time and position are related themselves, and
worst of all, the information one can get about positions and momenta is limited
by the laws of quantum mechanics (QM), just as it is the case in statistical mechanics.
 It is therefore intriguing to re-examine the concept of trajectories within the framework of quantum mechanics. 
 As it turns out, different formulations of QM have a very different interpretation of ``trajectories''.
 Nevertheless, independent of
 the formulation,
they all have to make sense of the fact that particle detectors  ``click'' at certain times and certain positions, which is the essence of the above definition of trajectory.
Typical interpretations of trajectories in different formulations of QM are the
\begin{itemize}
    \item Standard~\cite{aHeisenberg:1927zz,Schrodinger:1926gei} (e.g. Kopenhagen): A TJ has NO intrinsic meaning,
    it is only a result of a macroscopic statistical interpretation of the evolution of the wave-function $\psi(x,t)$.
    Whether and how this wave-function behaves under measurement (collapse, many worlds, ...) is subject of an additional discussion beyond our concern.
    \item
    Path integrals~\cite{Feynman:1948ur} (PI):
    TJ do ``exist'', but they are 
    neither determined by dynamical equations nor reduced to a single 
    path between the initial and final position of the particle.
    Instead, the weighted sum over all these random paths dictates the evolution of the wave-function $\psi(x,t)$.
    \item Stochastic mechanics~\cite{Nelson:1966sp,Kuipers:2023pzm,Kuipers:2023ibv} (SM):
    TJs do have intrinsic meaning, but
    the dynamics of all particles is governed by a stochastic differential equation. 
    \item 
    De Broglie-Bohm theory~\cite{Bohm:1951xw,Bohm:1951xx} (dBB):
    TJs do have intrinsic meaning. There exists a single path between the initial and final position, which is determined by an equation of motion. This equation of motion, however, is modified by the QM wave function $\psi$, called the pilot wave. The characteristic uncertainty of QM arises from the condition that the probability density is related to the statistical density of possible paths.
\end{itemize}
In this note we extend this list by a new item:
\begin{itemize}
    \item Earthquake quantization:
The TJs in the proposed EQQ share properties of both PI and dBB. On the one side 
a weighted sum over paths  dictates the evolution of a wave-function, like in PI. On the other side
these paths are determined by an equation of motion (similar to dBB).
%{\color{red}under instantaneous (gravitational) accelerations, that determine the sum over paths}.
This hybrid nature of the EQQ is shown in the schematic figure~\ref{fig_BMPIEQ}.
Whether this hybrid interpretation of trajectories can be linked to the many-worlds interpretation~\cite{Everett:1957hd}, where the universal wavefunction is objectively real, remains to be investigated.
\end{itemize}
%%%%%

%%%%%%%%%%%%%%%%%%%
\subsection{Construction of the non-relativistic path integral}

The PI of the non-relativistic point particle can be built from the infinitesimal propagator
\bea\nonumber
K_{(0)}(\vec x_i,0;\vec x_f,\Delta t)
&=&\left(\frac{m}{2 i \pi \hbar \Delta t}\right)^{3/2} 
\exp \left[\frac{im}{\hbar} \int_0^{\Delta t} dt'\frac{\dot{\vec x}^2}{2}\right] \\ \label{eq_K0}
&=& \left(\frac{m}{2 i \pi \hbar \Delta t}\right)^{3/2}  \exp \left[\frac{i}{\hbar} S^\ell_{if}(\Delta t)\right],
\eea
where the integral in the exponential is trivial, because the particle in this
infinitesimal construction travels on a straight line between the initial $\vec x_i$
and the final point $\vec x_f$
\be
S^\ell_{if}(\Delta t)=
\frac{m(\vec x_f-\vec x_i)^2}{2\Delta t}.
\ee
Note, that (\ref{eq_K0}) has a discretization ambiguity, which we fixed
by imposing time ordering $t_f>t_i$.
The transition to finite time intervals is obtained by the recursive use of the infinitesimal Kolmogorov relation
\be\label{eq_Kol0}
K_{(0)}(\vec x_i,0;\vec x_f,2\Delta t)=\int d^3x_1 K_{(0)}(\vec x_i,0;\vec x_1,\Delta t) K_{(0)}(\vec x_1,\Delta t;\vec x_f,2\Delta t).
\ee
This gives
\be\label{PI0}
\left(\frac{m}{2 i \pi \hbar t}\right)^{3/2}  \exp \left[\frac{i}{\hbar} S^\ell_{if}(t)\right]=
K(x_i,0;x_f,t)= 
\int \tilde {\mathcal D}x \exp \left[\frac{i}{\hbar}S\right],
\ee
where the action on the right hand side results from 
an integration along the ``drunken'' quantum path
\be\label{eq_S0int}
S=\int_0^{t} dt'\frac{m\dot{\vec x}^2(t')}{2}.
\ee
Note that in a strict mathematical sense, the integrals (\ref{PI0}) do not converge unless one defines a careful treatment of the infinities. For these more formal aspects we refer the reader to standard text books like~\cite{Albeverio:book}.

The measure in (\ref{PI0}) is defined as
\be\label{eq_Dx}
\int \tilde{\mathcal{D}}x\equiv
\left(\frac{m}{2i\hbar \pi \Delta t}\right)^{3/2}
\prod_{j=1}^N \left( \int d^3x_j
\left(\frac{m}{2i\pi \hbar \Delta t}\right)^{3/2} \right),
\ee
with $t_f-t_i=N \cdot \Delta t$.
This measure differs from the more common definition with only $d^3x$ due to the normalization factor
\be
\tilde {\mathcal{D}}x=
\left(\frac{m}{2i\pi \hbar \Delta t}\right)^{\frac{3(N+1)}{2}} 
{\mathcal{D}}x .
\ee
This definition is useful, since 
we want to express the right hand side of (\ref{PI0}) in terms of 
an exponential of the action~(\ref{eq_S0int}) without further normalizations factors.

%%%%%%%%%%%%%%%%%%%
\subsection{Metric and geodesics in the non-relativistic limit}

As outlined in the prologue, we are reinterpreting  random motion in terms of a random force and since gravity is the only known universal interaction, we expound
this random force as result of a geodesic motion in a random metric background.
For an arbitrary metric field $g_{\mu \nu}(x)$ we can define
deviations $\delta g_{\mu \nu}(x)$ from the flat Minkowski metric $\eta_{\mu \nu}$ as
\be\label{eq_metric}
g_{\mu \nu}(x)=\eta_{\mu \nu}+ \delta g_{\mu \nu}(x).
\ee
The motion of a point particle in such a background is given by the geodesic equation
\be\label{eq_Geodesic}
\frac{d^2 x^\mu}{ds^2}=-\Gamma^\mu_{\; \alpha \beta}\frac{dx^\alpha}{ds}\frac{dx^\beta}{ds}.
\ee
In the non-relativistic Newton limit this law of motion simplifies to
\be\label{eq_eomNR}
\ddot {x}^{j}=-
\Gamma^{j}_{\;00}\equiv
-\Gamma^{j}.
\ee

For shorter notation in our purely non-relativistic analysis we defined the relevant terms of the affine connection as a tri-dimensional vector 
$\vec \Gamma \equiv \hat{\mathbf{e}}_{j}\Gamma^{j}_{\;00} $ that plays the role of local and instantaneous force in the equation of motion~\eqref{eq_eomNR}.
This connection is given in terms of
derivatives of the metric
(\ref{eq_metric}) as
\be\label{eq_DefGamma}
\vec \Gamma= \vec \partial\frac{c^2}{2}\delta g_{00}(x).
\ee
Please note that, the connection $\Gamma^\mu_{\alpha \beta}$ is neither a vector, nor a tensor. Thus, using a vector arrow symbol in the definition (\ref{eq_DefGamma})
is only justified in the non-relativistic low curvature limit, where it transforms accordingly.

%%%%%%%%%%%%%%%
\subsection{Quantum and gravity}

A lot has been said, but little is yet understood about the interplay between gravity and the laws of quantum mechanics. Even more, in quantum-gravity, 
which is by itself a very hard topic~\cite{Deser:1999mh}, things get worse when one tries to include matter into the picture~\cite{tHooft:1974toh}.
There are numerous candidates for providing a valid theory of quantum gravity,
these formulations work with different fundamental degrees of freedom, such as the metric field $g_{\mu\nu}$ in the 2nd order formalism,
the connection $\Gamma^\mu_{\; \alpha \beta}$ in the 1st order formalism~\cite{Arnowitt:1962hi}, 
non-commutative or other geometries~\cite{Barrett:2015foa,Barrett:2019aig,Chamseddine:2022rnn,Reitz:2023ezz},
or a one-dimensional string (for a review see e.g. \cite{Carlip:2001wq}).

To the many open questions in this context we add another one:
{\it{
``Are geodesics in curved space-time truly classical, or do they only appear classical because we measure them in the classical limit?''}}
In the latter case, fluctuations of gravity could be seen and used as the actual cause of the path history in (\ref{PI0}),
in this note on the EQQ we will consider an integral over connections  $\Gamma$.

%%%%%%

%%%%%%%%%%%%%%%%%%%%%%%%%%%%%%%55
\section{Earthquake quantization}
%%%%%%%%%%%%%%%%%%%%%%%

Formally one can define the EQQ from
the functional integral (\ref{PI0}) and introducing an identity in terms of
a functional integral over the gravitational connection $\vec \Gamma$ and a functional delta function.
\be\label{eq_one}
\mathds{1} = \int {\mathcal{D}}\Gamma \delta \left[\vec \Gamma-{\textrm{eom}}(\vec x)\right].
\ee
Note that above, $\delta[f(\vec{x})]$ (with $[\cdot]$  brackets) refers to a functional delta function, while in the context of individual time-slizes it is actually a ``simple'' delta function (with $(\cdot)$ brackets).
The expression eom$(\vec x)$ 
stands for $-\ddot{\vec x}$.
The functional delta $\delta$ enforces the condition that at each time-step $\vec{x}=\vec{x}(t)$ is also a solution of the 
equation of motion (\ref{eq_eomNR}) for every given value of the connection $\vec{\Gamma}$, including appropriate boundary conditions.
Thus, a given $\vec x=\vec x_s(t)$ determines the connection $\vec \Gamma(t)$, or inversely a given connection determines a solution $\vec x_s$.
With this, (\ref{PI0}) can be rewritten as
\bea\label{eq_PIEQ}
K(\vec x_i,0;\vec x_f,t)&=&
\int \tilde {\mathcal D}x \cdot \mathds{1} \cdot \exp \left[\frac{i}{\hbar}S(x,\eta)\right]\\ \nonumber
&=&\int \tilde {\mathcal D}x \int {\mathcal{D}\Gamma} \cdot \delta\left[\vec \Gamma-{\textrm{eom}}(\vec x)\right] \exp \left[\frac{i}{\hbar}S(x,\eta)\right]\\
&=&\int
{\mathcal{D}\Gamma} J\exp \left[\frac{i}{\hbar}S(x,\eta)\right], \nonumber
\eea
where in the last step we exchanged the order of integrations, to explicitly perform the functional integral over the measure $\tilde{\mathcal{D}}x$ to obtain a Jacobian factor $J=\left(\frac{{\mathcal{D}} {\textrm{eom}}}{\tilde {\mathcal{D}}x}\right)^{-1}$.
Note that the exponentiated action is
a function of the trajectory $\vec x(t)$ in flat
Minkowski space-time $S=S(x,\eta)$. 
The connections of curved space-time $\vec \Gamma(t')$
only come into play when defining the trajectory.

%%%%%%%%%%%%%%%%%%%%%%%%%%%
\subsection{Free particle propagator}

To learn how the Jacobian with 
${\frac{{\mathcal{D}} {\textrm{eom}}}{\tilde {\mathcal{D}}x}}$ 
is calculated explicitly let's go back to the infinitesimal integral 
for two steps ($0\rightarrow \Delta t \rightarrow 2 \Delta t$), like in the
Kolmogorov relation (\ref{eq_Kol0}).

To generate the zig-zag motion of the path-integral construction from a potential, one needs to define a connection that is acting instantaneously at the intermediate time-steps.
For the two-step example the intermediate time is at $t=\Delta t$, and hence we define
\be
\vec \Gamma(x,t)\equiv \vec \gamma(x)\cdot \delta (t-\Delta t).
\ee
If the equations of motion (\ref{eq_eomNR}) are fulfilled,
we have
\be\label{eq_xddot}
\ddot {\vec x}=-\vec \gamma(x)\cdot \delta (t-\Delta t).
\ee
To average this equation over one step size, we integrate from $\Delta t/2$ to $3\Delta t/2$
and divide by $\Delta t$
\be\label{eq_xddoteom}
\ddot {\bar {\vec x}}=-\vec \gamma(t)\frac{1}{\Delta t}.
\ee
If the particle follows a PI trajectory with $\vec x_1$ as intermediate point,
the averaged acceleration is kinematically defined as
\be\label{eq_xddotkin}
\overline{\ddot { {\vec x}}}=\frac{1}{\Delta t}\left(\frac{\vec x_f-\vec x_1}{\Delta t}-\frac{\vec x_1-\vec x_i}{\Delta t}\right).
\ee
Identifying (\ref{eq_xddoteom}) with (\ref{eq_xddotkin}) allows to read-off the relation between $\vec x_1$ and the connection at $t=\Delta t$
\be\label{eq_eomPhi1}
 \vec \gamma_1=\frac{2}{\Delta t} \left( \vec x_1-\vec x^{\ell}_{\Delta t} \right).
\ee
Here
\be
\vec x^{\ell}_{\Delta t}=\frac{\vec x_i +\vec  x_f}{2},
\ee
is the position at $\Delta t$ that
corresponds to a straight undeflected line between initial $\vec x_i$ and final $\vec x_f$.
The infinitesimal Kolmogorov relation with the explicit Jacobian in 
(\ref{eq_PIEQ}) reads then
\be\label{eq_PIEQ2}
K(x_i,0;x_f,2 \Delta t)=
\int_{-\infty}^{+\infty} d^3\vec \gamma_1 \left(\frac{\Delta t}{2 }\right)^3
\left(\frac{m}{2i\pi \hbar \Delta t}\right)^{3(1+1)/2}
\exp \left[
\frac{i}{\hbar}
\left(S^\ell_{i1}(\Delta t)+S^\ell_{1f}(\Delta t)\right)
\right],
\ee
where $\vec x_1$ is the solution of (\ref{eq_eomPhi1})
\be\label{eq_x1ofg1}
\vec x_1= \vec x^{\ell}_{\Delta t} +\vec \gamma_1
\frac{\Delta t}{2}.
\ee
Note that due to (\ref{eq_x1ofg1}), the exponentiated action on the right hand side of (\ref{eq_PIEQ2}) is a function of $\vec x_i\, \vec x_f$, and $\vec \gamma_1$
\be\label{eq_S22}
\left(S^\ell_{i1}(\Delta t)+S^\ell_{1f}(\Delta t)\right)= S(\vec x_i\, \vec x_f, \vec \gamma_1,\Delta t)=m
\frac{(\vec x_f-\vec x_i)^2+\vec \gamma_1^2 \Delta t^2}{4 \Delta t},
\ee
which has its minimum at $\vec \gamma_1=0$.
For finite time slicing $\Delta t\rightarrow t/2$, this relation allows to factorize the 
exponential of the action of a straight line connecting $\vec x_i$ and $\vec x_f$
\be\label{eq_PIEQ3}
\exp \left[\frac{i}{\hbar} S_{if}^\ell(t)\right]=
\left( \frac{ tm }{2 i \pi \hbar}\right)^{3/2}
\int_{-\infty}^{+\infty} d^3\vec 
\gamma_1
\exp \left[\frac{i}{\hbar} S_{i1}^\ell(t/2)\right]\cdot 
\exp \left[\frac{i}{\hbar} S_{1f}^\ell(t/2)\right].
\ee
Now, one can subdivide all steps into two parts and iteratively insert the prescription (\ref{eq_PIEQ3}) into itself.
Repeating this procedure n-times allows to construct
a path integral with an even number of steps $2^n$ and an odd number of intermediate integrals
$N=2^n-1$.
One finds
\be\label{eq_PIEQ4}
\exp \left[\frac{i}{\hbar} S_{if}^\ell(t)\right]=\int
\tilde {\mathcal{D}}\gamma \exp \left(\frac{i}{\hbar}S\right),
\ee
where the action integral on the right hand side is only a function of the initial and final position, the total time and the intermediate connections
$S=S(\vec x_i, \vec x_f,\vec \gamma_k,t)$ (\ref{eq_S22}), resulting
from the recursive insertion of
\be
\vec x_k=\frac{\vec x_{k-1}+\vec x_{k+1}}{2}+\vec \gamma_k\frac{\Delta t}{2}
\ee
into the discretized action.
Further, the
normalized integral measure is defined as
\be\label{eq_DPhi}
\int \tilde {\mathcal{D}}\gamma\equiv
%(N+1)^{-(1+\ln (N-1))}
\left(
\frac{tm}{2 i \pi \hbar}
\right)^{\frac{3}{2}N}
\prod_{k=1}^N 
\left( \int d^3 \vec \gamma_k
\right).
\ee
The propagator corresponding to the path integral
(\ref{eq_PIEQ4}) is obtained by a multiplication with
the normalization defined in (\ref{eq_K0}).

%%%%%%%%%%%%%%%%%%%%%%%%%%%%%%%%%%%%%%%
\section{Continuum formulation}
The previous procedure can also be formulated directly in the continuum limit. We shall consider two examples that can be explicitly solved exactly, in order to illustrate in detail the applicability of the EQQ procedure: (A) the free particle case, and (B) the harmonic oscillator.

%%%%%%%%%%%%%%%%%%%%%%%%%%%%%%
\subsection{Free particle}
Let us consider the path-integral  representation of the propagator for the free particle
\bea
K(\vec{x}_i,t_i;\vec{x}_f,t_f) = \int_{x(t_i)=x_i}^{x(t_f)=x_f} \mathcal{D}x\, \exp{\left[\frac{i}{\hbar}S[x]\right]},
\label{eq_Kcont}
\eea
where the action in this case is defined by the integral
\bea
S[x] = \int_{t_i}^{t_f}\frac{m}{2}\left[\dot{\vec{x}}(t)\right]^2
\label{eq_Sfree}
\eea
In the continuum representation \eqref{eq_Kcont}, the functional integral includes all trajectories that satisfy the specified boundary conditions, i.e.
\be
\vec{x}(t=t_i) = \vec{x}_i,\,\,\,\,\,\, \vec{x}(t=t_f) = \vec{x}_f.
\label{eq_BC}
\ee
Among those paths, we identify as usual the "classical path" as the one satisfying the variational stationary action principle $\delta S = 0$, subject to fixed boundary conditions Eq.~\eqref{eq_BC}, i.e. $\delta \vec{x}(0) = \delta\vec{x}(t_f) =0$,
\bea
\delta S = \int_{t_i}^{t_f}\left\{ \frac{d}{dt}\left( m\dot{\vec{x}}\cdot\delta\vec{x} \right) - m \ddot{\vec{x}}\cdot\delta\vec{x} \right\}dt = -\int_{t_i}^{t_f}m \ddot{\vec{x}}\cdot\delta\vec{x}\, dt = 0
\eea
Therefore, the "classical path" is specified by the solution to the differential equation
\be
-m\ddot{\vec{x}}_{cl}(t) =0,
\label{eom_class}
\ee
subject to the boundary conditions Eq.~\eqref{eq_BC}. Trivially, this solution is given by the function
\be
\vec{x}_{cl}(t) = \vec{x}_i + \frac{t - t_i}{t_f - t_i} \left( \vec{x}_f - \vec{x}_i\right),
\label{eq_xclass}
\ee
that represents a trajectory with a {\bf{constant velocity}} determined by the expression
\be
\dot{\vec{x}}_{cl}(t) \equiv \vec{v}_{cl} = \frac{\left( \vec{x}_f - \vec{x}_i\right)}{t_f - t_i}
\ee
that connects the initial and final points of the finite propagator.

Quantum mechanics introduces random fluctuations with respect to such a straight classical path. One may wonder about the physical interpretation of the origin of such fluctuations. Following our previous analysis, one may attribute their source to a random fluctuating metric, whose effect is equivalent to a random force $\vec{\Gamma}(t)$, such that the evolution of the trajectory is a dynamical equation that incorporates such force,
\be
- m \ddot{\vec{x}}(t) \equiv eom[\vec{x}(t)] = \vec{\Gamma}(t).
\label{eq_random_dyn}
\ee
Given the specified boundary conditions in Eq.~\eqref{eq_BC}, the solution to this differential equation is given by a linear combination of the solution to the homogeneous problem Eq.~\eqref{eom_class} (which is nothing but the classical trajectory according to Eq.~\eqref{eq_xclass}) and a convolution between the Green's function for the linear operator and the impulsive source,
\bea
\vec{x}(t) = \vec{x}_{cl}(t) + \int_{t_i}^{t_f}dt' \hat{g}_0(t,t') \vec{\Gamma}(t').
\label{eq_x_sol}
\eea
For the Green's function, we may alternatively use the formal notation
\be
\hat{g}_0(t,t') = \mathbf{1}\left[-m\frac{d^2}{dt^2} \right]^{-1}\delta(t-t').
\label{eq_g_inv}
\ee
Therefore, the mathematical solution to Eq.~\eqref{eq_random_dyn} may be interpreted such that any arbitrary path $\vec{x}(t)$ is defined by a superposition of the classical path and an instantaneous fluctuation $\vec{x}_Q(t)$, i.e.
\be
\vec{x}(t) = \vec{x}_{cl}(t) + \vec{x}_Q(t),
\label{eq_xQ}
\ee
where the fluctuation $\vec{x}_Q(t)$ is given explicitly by Eq.~\eqref{eq_x_sol} in terms of the random force
\be
\vec{x}_{Q}(t) = \int_{t_i}^{t_f}dt' \hat{g}_0(t,t') \vec{\Gamma}(t')
\label{eq_xQ_sol}
\ee
Clearly, by definition and given the fixed boundary conditions established by Eq.~\eqref{eq_BC}, such fluctuations must vanish at the two ends of the path, i.e.
\be
\vec{x}_Q(t_i) = 0,\,\,\,\,\vec{x}_Q(t_f) = 0.
\label{eq_BCQ}
\ee
 Taking the first and second time-derivative of Eq.~\eqref{eq_xQ}, we have the subsequent relations
\bea
\dot{\vec{x}}(t) &=&  \dot{\vec{x}}_{cl}(t) + \dot{\vec{x}}_Q(t) \nonumber\\
-m\ddot{\vec{x}}(t) &=&   -m\ddot{\vec{x}}_Q(t) = \vec{\Gamma}(t),
\label{eq_xQdyn}
\eea
where we applied the classical equation of motion $\ddot{\vec{x}}_{cl}(t)=0$.

Inserting the definition Eq.~\eqref{eq_xQ} into the action Eq.~\eqref{eq_Sfree}, we trivially obtain
\bea
S[x] &=& \int_{t_i}^{t_f} dt \frac{m}{2}\left[ \dot{\vec{x}}_{cl}(t) + \dot{\vec{x}}_{Q}(t) \right]^2 = \frac{m}{2}\vec{v}_{cl}^2(t_f - t_i) + \frac{m}{2}\vec{v}_{cl}\cdot\int_{t_i}^{t_f}\dot{\vec{x}}_{Q}(t)dt
+ \int_{t_i}^{t_f}\frac{m}{2}\left[\dot{\vec{x}}_{Q}(t)\right]^2\,dt\nonumber\\
&=& \frac{m}{2}\frac{\left(\vec{x}_f - \vec{x}_i\right)^2}{t_f-t_i} - \int_{t_i}^{t_f}\frac{m}{2}\vec{x}_{Q}(t)\cdot\ddot{\vec{x}}_{Q}(t)\,dt,
\eea
where the second term exactly vanishes given the boundary conditions Eq.~\eqref{eq_BCQ} for the fluctuations, and we integrated by parts to transform the third integral. Therefore, we have extracted the "classical path" contribution to the action, and the remaining part is pure "quantum fluctuations". The splitting into "classical" (i.e. deterministic) coordinates and "quantum" fluctuations defined by Eq.~\eqref{eq_xQ} induces a corresponding redefinition of the functional measure $\mathcal{D}x = \mathcal{D}x_Q$, such that the propagator in Eq.~\eqref{eq_Kcont} becomes
\bea
K(\vec{x}_i,t_i;\vec{x}_f,t_f) = e^{\frac{i m \left( \vec{x}_f - \vec{x}_i \right)^2}{2\hbar(t_f - t_i)}}\int_{x_Q(t_i)=0}^{x_Q(t_f)=0} \mathcal{D}x_Q\, \exp{\left[-\frac{i}{2\hbar} \int_{t_i}^{t_f}m\vec{x}_{Q}(t)\cdot\ddot{\vec{x}}_{Q}(t)\,dt\right]}.
\label{eq_Kfluct}
\eea
We can now insert into Eq.~\eqref{eq_Kfluct} the functional identity that enforces the dynamics of the random fluctuations as defined by Eq.~\eqref{eq_xQdyn},
\bea
\int \mathcal{D}\Gamma \delta\left[ -m\ddot{\vec{x}}_Q(t) - \vec{\Gamma}(t) \right] = 1.
\eea
Therefore, we have
\bea
K(\vec{x}_i,t_i;\vec{x}_f,t_f) &=& e^{\frac{i m \left( \vec{x}_f - \vec{x}_i \right)^2}{2\hbar(t_f - t_i)}}\int \mathcal{D}\Gamma \int_{x_Q(t_i)=0}^{x_Q(t_f)=0} \mathcal{D}x_Q\, \delta\left[ -m\ddot{\vec{x}}_Q(t) - \vec{\Gamma}(t) \right]\exp{\left[-\frac{i}{2\hbar} \int_{t_i}^{t_f}m\vec{x}_{Q}(t)\cdot\ddot{\vec{x}}_{Q}(t)\,dt\right]}\nonumber\\
&=& e^{\frac{i m \left( \vec{x}_f - \vec{x}_i \right)^2}{2\hbar(t_f - t_i)}}\int \mathcal{D}\Gamma
\exp{\left[\frac{i}{2\hbar} \int_{t_i}^{t_f}dt\int_{t_i}^{t_f}dt'\,\vec{\Gamma}(t)\hat{g}_{0}(t,t')\vec{\Gamma}(t')\right]}
\int_{x_Q(t_i)=0}^{x_Q(t_f)=0} \mathcal{D}x_Q\, \delta\left[ -m\ddot{\vec{x}}_Q(t) - \vec{\Gamma}(t) \right],\nonumber\\
\label{eq_K_Gam}
\eea
where in the second line we used the functional delta to express the explicit solution for $\vec{x}_Q(t)$ after Eq.~\eqref{eq_xQ_sol}.
We notice, as before, that the second integral reduces to a functional determinant of the differential operator that defines the equation of motion $eom[\vec{x}_Q] = -m\ddot{\vec{x}}_Q(t)$, i.e.
\bea
\int_{x_Q(t_i)=0}^{x_Q(t_f)=0}\mathcal{D}x_Q\, \delta\left[ -m\ddot{\vec{x}}_Q(t) - \vec{\Gamma}(t) \right] = \left[\frac{\mathcal{D}eom[\vec{x}_Q]}{\mathcal{D}\vec{x}_Q} \right]^{-1} = \left({\rm{Det}}\left[ -\mathbf{1}m\frac{d^2}{dt^2} \right]\right)^{-1},
\eea
which (calculated explicitly in Appendix A) turns out to be independent of the random force $\vec{\Gamma}(t)$, and hence can be factored out. Therefore, we arrive at the desired conceptual form of the propagator,
\be
K(\vec{x}_i,t_i;\vec{x}_f,t_f) = e^{\frac{i m \left( \vec{x}_f - \vec{x}_i \right)^2}{2\hbar(t_f - t_i)}}
\left[\frac{\mathcal{D}eom[\vec{x}_Q]}{\mathcal{D}\vec{x}_Q} \right]^{-1}  \int \mathcal{D}\Gamma
\exp{\left[\frac{i}{2\hbar} \int_{t_i}^{t_f}dt\int_{t_i}^{t_f}dt'\,\vec{\Gamma}(t)\hat{g}_0(t,t')\vec{\Gamma}(t')\right]},
\ee
where clearly there is an explicit overall phase given by the action at the classical trajectory, and an additional amplitude depending on a gaussian path-integral over the configurations of the random force, with a kernel $\hat{g}_0(t-t')$ corresponding to the Green's function for the operator that defines the time-evolution under such a random force, i.e. Eq.~\eqref{eq_g_inv}.

We can complete the analysis of this case, by calculating the remaining Gaussian functional integral over the random force $\vec{\Gamma}(t)$, that reduces to
\bea
\int \mathcal{D}\Gamma
\exp{\left[\frac{i}{2\hbar} \int_{t_i}^{t_f}dt\int_{t_i}^{t_f}dt'\,\vec{\Gamma}(t)\hat{g}_0(t,t')\vec{\Gamma}(t')\right]} = \mathcal{N}(t_f-t_i) \left({\rm{Det}}\left[ -\mathbf{1}m\frac{d^2}{dt^2} \right]\right)^{1/2},
\eea
where $\mathcal{N}(t_f-t_i)$ is the normalization factor for the path-integral measure. Therefore, the final result in Eq.~\eqref{eq_Kfluct} is
\bea
K(\vec{x}_i,t_i;\vec{x}_f,t_f) &=& e^{\frac{i m \left( \vec{x}_f - \vec{x}_i \right)^2}{2\hbar(t_f - t_i)}} \mathcal{N}(t_f-t_i) \left({\rm{Det}}\left[ -\mathbf{1}m\frac{d^2}{dt^2} \right]\right)^{-1/2}.
\eea
Finally, the normalization factor is determined by the condition
\be
\mathcal{N}(t_f-t_i) \left({\rm{Det}}\left[ -\mathbf{1}m\frac{d^2}{dt^2} \right]\right)^{-1/2} = \left( \frac{m}{2i\pi\hbar(t_f - t_i)} \right)^{3/2}.
\label{eq_Knorm}
\ee
This expression, that fixes the normalization factor, will be applied in our next explicit example.

%%%%%%%%%%%%%%%%%%%%%%%%%%%%%%%%%%%%
\subsection{The harmonic potential}
It is interesting to check that the procedure presented in the previous section, including the normalization in Eq.~\eqref{eq_Knorm}, can be correctly applied to reproduce the propagator for another well known example, the case of a particle submitted to a harmonic potential. In this case, we have that the action is given by
\be
S[x] = \int_{t_i}^{t_f}\left\{ \frac{m}{2}\left[\dot{\vec{x}}(t)\right]^2 - \frac{m\Omega^2}{2}\left[ \vec{x}(t) \right]^2 \right\}dt,
\ee
with specified boundary conditions as in Eq.~\eqref{eq_BC}. The classical path, in this case, follows again from the variational principle,
\bea
\delta S = \int_{t_i}^{t_f}\left\{ \frac{d}{dt}\left( m\dot{\vec{x}}\cdot\delta\vec{x} \right) - m \ddot{\vec{x}}\cdot\delta\vec{x} - m \Omega^2\vec{x}(t)\cdot\delta\vec{x} \right\}dt = -\int_{t_i}^{t_f}\left(m \ddot{\vec{x}} + m\Omega^2 \vec{x}(t)\right)\cdot\delta\vec{x}\, dt = 0.
\eea
The "classical path" is then a solution to the deterministic, homogeneous differential equation
\bea
-m\ddot{\vec{x}}_{cl} - m\Omega^2 \vec{x}_{cl} = 0,
\label{eq_harm_eom_class}
\eea
subject to the boundary conditions Eq.~\eqref{eq_BC}. The explicit analytical solution is then given by the function
\bea
\vec{x}_{cl}(t) = \frac{\vec{x}_i\sin\left(\Omega(t_f - t) \right) + \vec{x}_f \sin\left(\Omega(t - t_i) \right)}{\sin\left( \Omega (t_f - t_i)  \right)}.
\label{eq_xcl_harm}
\eea
As in our previous example, we may attribute the existence of fluctuations to an arbitrary path with respect to the classical trajectory, due to the presence of a random force (fluctuations in the metric) $\vec{\Gamma}(t)$, such that the actual trajectories are the solutions to the inhomogeneous differential equation
\bea
-m\ddot{\vec{x}}(t) - m\Omega^2\vec{x}(t) = \vec{\Gamma}(t).
\label{eq_harm_eom_xq}
\eea
The solution to this inhomogeneous differential equation is given by a linear combination of the solution to the homogeneous problem (again, just the classical Eq.~\eqref{eq_xcl_harm}) and a convolution of the impulsive force with the Green's function of the equation of motion
\bea
\vec{x}(t) = \vec{x}_{cl}(t) + \int_{t_i}^{t_f}dt'\hat{g}_{\Omega}(t,t')\vec{\Gamma}(t') \equiv \vec{x}_{cl}(t) + \vec{x}_Q(t),
\eea
where in this case we have the formal definition
\bea
\hat{g}_{\Omega}(t,t') = \mathbf{1}\left[-m\frac{d^2}{dt^2} - m\Omega^2 \right]^{-1}\delta(t-t').
\label{eq_ginv_harm}
\eea
Let us now consider an actual path, which as stated before may be interpreted as a superposition of the "classical path" and the random fluctuation, $\vec{x}(t) = \vec{x}_{cl}(t) + \vec{x}_{Q}(t)$, such that
\bea
\dot{\vec{x}}(t) &=& \dot{\vec{x}}_{cl}(t) + \dot{\vec{x}}_Q(t) \nonumber\\
-m\ddot{\vec{x}}(t) - m\Omega^2\vec{x}(t) &=& -m\ddot{\vec{x}}_{Q}(t) -m \Omega^2\vec{x}_{Q}(t) = \vec{\Gamma}(t).
\eea
Evaluating the action for any such trajectories, after some algebra we have
\bea
S[x] = S[\vec{x}_{cl}(t) + \vec{x}_Q(t)] = S[\vec{x}_{cl}(t)]
+ \frac{1}{2}\int_{t_i}^{t_f}dt\,\vec{x}_{Q}(t)\cdot\left(  -m\ddot{\vec{x}}_{Q}(t) - m\Omega^2\vec{x}_Q(t)\right),
\eea
where we obtained the action for the classical path
\be
S[\vec{x}_{cl}(t)] = \frac{m \Omega}{2}\left[ 
\left( \vec{x}_f^2 + \vec{x}_i^2 \right)\cot\left(\Omega(t_f - t_i) \right)
-2\frac{\vec{x}_i\cdot\vec{x}_f}{\sin\left( \Omega (t_f - t_i) \right)}
\right]
\ee
Therefore, as in our previous example, the path-integral representation for the propagator becomes
\be
K(\vec{x}_i,t_i;\vec{x}_f,t_f) = e^{\frac{i}{\hbar}  S[\vec{x}_{cl}(t)]} \int_{x_Q(t_i)=0}^{x_Q(t_f)=0}\mathcal{D}x_Q\, \exp{\left[\frac{i}{2\hbar}\int_{t_i}^{t_f}dt\,\vec{x}_{Q}(t)\cdot\left(  -m\ddot{\vec{x}}_{Q}(t) - m\Omega^2\vec{x}_Q(t)\right)\right]}.
\label{eq_Kfluct_harm}
\ee
Now, we can introduce the explicit dynamics for the fluctuations $\vec{x}_Q(t)$ via the functional integral identity
\be
\int\mathcal{D}\Gamma \delta\left[ -m\ddot{\vec{x}}_Q(t) - m\Omega^2\vec{x}_Q(t) - \vec{\Gamma}(t) \right] = 1,
\ee
such that after similar manipulations as in the previous case, we obtain
\bea
K(\vec{x}_i,t_i;\vec{x}_f,t_f) &=& e^{\frac{i}{\hbar}  S[\vec{x}_{cl}(t)]} \int\mathcal{D}\Gamma e^{  
\frac{i}{2\hbar}\int_{t_i}^{t_f}dt\int_{t_i}^{t_f}dt'\,\vec{\Gamma}(t)\hat{g}_{\Omega}(t,t')\vec{\Gamma}(t')
}\int_{x_Q(t_i)=0}^{x_Q(t_f)=0}\mathcal{D}x_Q\, \delta\left[ -m\ddot{\vec{x}}_Q(t) - m\Omega^2\vec{x}_Q(t) - \vec{\Gamma}(t) \right]\nonumber\\
&=& e^{\frac{i}{\hbar}  S[\vec{x}_{cl}(t)]}
\left[\frac{\mathcal{D}eom[\vec{x}_Q]}{\mathcal{D}\vec{x}_Q} \right]^{-1}  \int \mathcal{D}\Gamma
\exp{\left[\frac{i}{2\hbar} \int_{t_i}^{t_f}dt\int_{t_i}^{t_f}dt'\,\vec{\Gamma}(t)\hat{g}_{\Omega}(t,t')\vec{\Gamma}(t')\right]}.
\label{eq_KGam_harm}
\eea
Here, for $eom[x_Q(t)] \equiv -m\ddot{\vec{x}}_Q(t) - m\Omega^2\vec{x}_Q(t)$, we applied the explicit result
\bea
\int_{x_Q(t_i)=0}^{x_Q(t_f)=0}\mathcal{D}x_Q\, \delta\left[ -m\ddot{\vec{x}}_Q(t) - m\Omega^2\vec{x}_Q(t) - \vec{\Gamma}(t) \right] = \left[\frac{\mathcal{D}eom[\vec{x}_Q]}{\mathcal{D}\vec{x}_Q} \right]^{-1} = \left({\rm{Det}}\left[\mathbf{1}\left( -m\frac{d^2}{dt^2} - m\Omega^2\right)\right]\right)^{-1}.
\label{eq_det_harm_op}
\eea
Clearly, the representation Eq.~\eqref{eq_KGam_harm} for the propagator represents an analogous conceptual form as in the free-particle case, i.e. a phase given by the classical action, and an amplitude expressed as a Gaussian functional integral over the random force $\vec{\Gamma}(t)$, whose kernel is now provided by the corresponding Green's function in Eq.~\eqref{eq_ginv_harm}.

Let us now complete this example, by evaluating the remaining Gaussian integral over the noise, to obtain (using the formal definition Eq.~\eqref{eq_ginv_harm} of the Green's function),
\bea
\int \mathcal{D}\Gamma
\exp{\left[\frac{i}{2\hbar} \int_{t_i}^{t_f}dt\int_{t_i}^{t_f}dt'\,\vec{\Gamma}(t)\hat{g}_{\Omega}(t,t')\vec{\Gamma}(t')\right]} = \mathcal{N}(t_f-t_i) \left({\rm{Det}}\left[ \mathbf{1}\left(-m\frac{d^2}{dt^2} - m\Omega^2\right)  \right]\right)^{1/2},
\eea
with the same overall normalization factor defined in Eq.~\eqref{eq_Knorm}. Therefore, substituting these results, the final explicit expression for the propagator is
\bea
K(\vec{x}_i,t_i;\vec{x}_f,t_f) &=& e^{\frac{i}{\hbar}  S[\vec{x}_{cl}(t)]} \mathcal{N}(t_f-t_i) \left({\rm{Det}}\left[ \mathbf{1}\left(-m\frac{d^2}{dt^2} - m\Omega^2\right)  \right]\right)^{-1/2}\nonumber\\
&=& e^{\frac{i}{\hbar}  S[\vec{x}_{cl}(t)]} \left( \frac{m}{2i\pi\hbar(t_f - t_i)} \right)^{3/2} \frac{\left({\rm{Det}}\left[ \mathbf{1}\left(-m\frac{d^2}{dt^2} - m\Omega^2\right)  \right]\right)^{-1/2}}{\left({\rm{Det}}\left[ \mathbf{1}\left(-m\frac{d^2}{dt^2} \right)  \right]\right)^{-1/2}}
\eea
As shown in detail in the Appendix A, the ratio between the functional determinants is given by
\be
\frac{\left({\rm{Det}}\left[ \mathbf{1}\left(-m\frac{d^2}{dt^2} - m\Omega^2\right)  \right]\right)^{-1/2}}{\left({\rm{Det}}\left[ \mathbf{1}\left(-m\frac{d^2}{dt^2} \right)  \right]\right)^{-1/2}} = \left( \frac{\sin\left( \Omega(t_f - t_i)\right)}{\Omega (t_f - t_i)} \right)^{-3/2},
\ee
and hence the propagator finally reduces to the exact (and well-known) expression
\be
K(\vec{x}_i,t_i;\vec{x}_f,t_f) = \left( \frac{m\Omega}{2\pi\,i\hbar \sin\left( \Omega(t_f - t_i)\right) }\right)^{3/2} \exp\left[ \frac{i\,m \Omega}{2\hbar}\left[ 
\left( \vec{x}_f^2 + \vec{x}_i^2 \right)\cot\left(\Omega(t_f - t_i) \right)
-2\frac{\vec{x}_i\cdot\vec{x}_f}{\sin\left( \Omega (t_f - t_i) \right)}\right]\right]
\ee

%%%%%%%%%%%%%%%%%%%%%%%%%%%
\subsection{Propagator for generic potentials}

A direct solution of the path integral 
with external potentials is typically only possible in special cases, such as the harmonic oscillator.
For the other cases, one has to recur to perturbation theory. Here, we  will follow the notation of the classic book~\cite{Feynman:1965}.

To show that EQQ is compatible with generic (weak) potentials we start of with the standard definition of the Kernel
\be\label{eq_PIV}
K_V(\vec x_i,t_i;\vec x_f,t_f)= \int {\mathcal D}x \exp \left[\frac{i}{\hbar} \int_{t_i}^{t_f} dt'
\left\{m\frac{\dot {\vec x}^2}{2}-V(\vec x,t')\right\}\right].
\ee
Note that we take the potential as a result of an external non-gravitational interaction, but it could also arise if one expands (\ref{eq_metric}) around a  non-flat background metric $g^{bg}_{\mu \nu}$ instead of the flat $\eta_{\mu \nu}$.
For
sufficiently small potentials, we expand the exponential containing the potential
\be\label{eq_ExpandExpV}
\exp \left[ -\frac{i}{\hbar}\int_{t_i}^{t_f} ds V(\vec x,s)\right]=
1- \frac{i}{\hbar}\int_{t_i}^{t_f} ds V(\vec x,s) +
\frac{1}{2\hbar^2} \left( \int_{t_i}^{t_f} ds V(\vec x,s)\right)^2 + \dots
\ee
Inserting this back into (\ref{eq_PIV})
we can write the propagator as
\be\label{eq_K0K1K2}
K_V(\vec x_i,t_i;\vec x_f,t_f)= K_0(\vec x_i,t_i;\vec x_f,t_f)+
K^{(1)}(\vec x_i,t_i;\vec x_f,t_f)+K^{(2)}(\vec x_i,t_i;\vec x_f,t_f)+ \dots .
\ee
The first term in this series expansion is given by (\ref{PI0}). Let's now illustrate
how the following terms are treated by examining $K^{(1)}$.
By commuting the integration order between
$ds \leftrightarrow dx_i$ we can write
\be
K^{(1)}(\vec x_i,t_i;\vec x_f,t_f)= - \frac{i}{\hbar}\int_{t_i}^{t_f} ds
\; F_s
\ee
with
\be
F_s= \int {\mathcal{D}}x \exp \left[
\frac{im}{\hbar} \int_{t_{i}}^{t_{f}} dt'\frac{\dot {\vec x}^2}{2}\right]\cdot V\left(\vec x(s),s\right).
\ee
Without loss of generality, and for the sake of notational simplicity, from now on we set $t_i = 0$ and $t_f \equiv t$.  Now we
discretize the time intervals
and realize that all time steps before $t'=s$ are just an integration of the exponential of the free particle,
just like all integrals after $t'=s$.
For these, we can thus use the EQQ path integral~(\ref{eq_PIEQ2}).
The remaining position integral is
\be\label{eq_Fs}
F_s= \int d^3x_s \left(\frac{-4 m^2}{\pi^2 \hbar ^2s (t-s)}\right)^{\frac{3}{2}} 
\exp \left[\frac{im}{2\hbar}\left(\frac{(\vec x_s-\vec x_i)^2}{s} +\frac{(\vec x_f-\vec x_s)^2}{t-s}\right)\right] 
V\left(\vec x_s\right).
\ee
The exponentials only contain straight lines with two different velocities
before and after the transition time $s$.
This velocity-change can, in analogy to the previous section,
be calculated from a PI perspective (\ref{eq_xddotkin})
and from the
perspective of the equations of motion (\ref{eq_xddoteom}).
Now, the difference is that the time variable $s$ is not right in the middle of
the total time $t$. Thus, the PI definition of average acceleration becomes
\be\label{eq_xddotkins}
\ddot {\bar {\vec x}}=\frac{2}{ t}\left(\frac{\vec x_f-\vec x_s}{t-s}-\frac{\vec x_s-\vec x_i}{s}\right),
\ee
and the definition according to the equations of motion can be obtained
from integrating (\ref{eq_xddot}) from $s/2$ to $s+(t-s)/2$ and dividing $t/2$
\be\label{eq_xddoteoms}
\ddot {\bar {\vec x}}=-\vec \gamma_s\frac{2}{ t}.
\ee
Comparing  (\ref{eq_xddotkins}) with (\ref{eq_xddoteoms}) yields 
the needed relation between $\vec x_s$ and $\vec \gamma_s$
\be\label{eq_gammas}
\vec \gamma_s=\frac{t}{s (t-s)}\left(
\vec x_s-\vec x^{\ell}_s
\right).
\ee
This is the generalization of the previous relation (\ref{eq_eomPhi1}), now for the case of different lengths of the time steps of  the foliation.
Thus, with the help of (\ref{eq_gammas}) we can write the integrand of (\ref{eq_Fs}) purely in terms of ($\vec x_i$, $\vec x_f$, and $\vec \gamma_s$) giving
\be
F_s= \int d^3\gamma_s \left(\frac{-4 m^2s (t-s)}{\pi^2 \hbar ^2t^2}\right)^{\frac{3}{2}} 
\exp \left[\frac{im}{2\hbar}\left(\frac{(\vec x_s-\vec x_i)^2}{s} +\frac{(\vec x_f-\vec x_s)^2}{t-s}\right)\right] 
V\left(\vec x_s\right).
\ee
The above steps and definitions can now be applied to all following terms in the 
expansion of the propagator (\ref{eq_K0K1K2}), providing 
an EQQ prescription for arbitrary potentials.

%%%%%%%%%%%%%%%%%%%%%%%%%%%
\subsection{Schr\"odinger equation}

The propagators (\ref{eq_PIV})
can be used to determine the evolution of a wave function
\bea\label{eq_psitp1}
 \psi(\vec x_f,t+\delta t)&=&
\int d^3x_i K_V(x_i,t;x_f,t+\delta t) \psi(\vec x_i,t)\\ \nonumber
&= &\int d^3x_i
\left( \frac{m}{2i \pi \hbar \delta t}\right)^{3/2}
\exp \left( \frac{i}{\hbar}m \frac{(\vec x_f-\vec x_i)^2}{2 \delta t}\right)
\exp \left( -\frac{i}{\hbar} V\left(\frac{\vec x_f +\vec x_i}{2}\right)\delta t\right)
 \psi(\vec x_i,t).
\eea
Now, we introduce 
the EQQ connections  at the initial time
\be
\vec \Gamma (t')= \vec \gamma_i \delta (t'-t).
\ee
These produce the velocity kicks,
which determine the relation
between initial and final velocities and positions
\bea
\vec v_f &=& \vec v_i - \vec \gamma_i\\ 
\vec x_f &=& \vec x_i + \vec v_f \delta t. 
\eea
With this, we can change the $d^3 x_i$ integration to a $d^3 \gamma_i$ integration, yielding 
\bea\label{eq_psitp2}
 \psi(\vec x_f,t+\delta t)&=&
\int d^3\gamma_i
\left( \frac{m\delta t}{2i \pi \hbar }\right)^{3/2}\\ \nonumber
&&
\exp \left( \frac{i}{\hbar}m (\vec v_i-\vec \gamma_i)^2\delta t\right)
\exp \left( -\frac{i}{\hbar} V\left(\vec x_f-\frac{\vec v_i-\vec \gamma_i}{2}\delta t\right)\delta t\right)
 \psi\left(\vec x_f-(\vec v_i-\vec \gamma_i)\delta t,t\right).
\eea
This is 
the EQ version of relation (\ref{eq_psitp1}). Interestingly,
it seems to suggest that
the evolved wave function at $t+\delta t$  depends on the velocity of the paths at $t$. 
This dependence is, however, fictitious. To see this,
we will now use the integral
relation~(\ref{eq_psitp2}) for the evolution of wave functions to derive a differential eqution with the same purpose, the Schr\"odinger equation. 
For this, we adapt the usual steps
outlined in~\cite{Feynman:1965}.
First, we replace $d^3\gamma$ by a new integration variable
\be
\vec \omega= (\vec v_i-\vec \gamma_i)\delta t.
\ee
This simplifies the Gaussian integral.
Second, we expand one of the exponentials for small potentials and small $\delta t$
\be
\exp \left( -\frac{i}{\hbar} V\left(\vec x_f-\frac{\vec v_i-\vec \gamma_i}{2}\delta t\right)\delta t\right)\approx 1 -\frac{i}{\hbar} V(\vec x_f)\delta t.
\ee
Third, we expand the wave function for small $\delta t$
\be
\tilde \psi\left(\vec x_f-(\vec v_i-\vec \gamma_i)\delta t,t\right)\approx
\tilde \psi\left(\vec x_f,t\right)-\vec \omega
\cdot \vec \nabla
\tilde \psi\left(\vec x_f,t\right)
+\frac{1}{2}\omega^j \omega^k  \nabla_j \nabla_k\tilde \psi\left(\vec x_f,t\right).
\ee
After performing the integral in $d^3\omega = (\delta t)^3 d^3\gamma_i$, (\ref{eq_psitp2}) reads
\bea\label{eq_psitp3}
\psi(\vec x_f,t+\delta t)&=&
\left( 
1
-i V(\vec x_f)\frac{ \delta t}{\hbar}
+i\frac{\hbar \delta t}{2m}\vec \nabla^2
\right)
\psi\left(\vec x_f,t\right)
\eea
Then we pull
$\tilde \psi(\vec x_f,t)$ to the left, and multiply by $i \hbar / \delta t$. This gives, in the limit of $\delta t\rightarrow 0$, the familiar Schr\"odinger equation
\be
i \hbar\partial_t\psi(\vec x_f,t)
=
\left(-\frac{\hbar^2}{2m}\vec \nabla^2+V(\vec x_f) \right) \psi(\vec x_f,t).
\ee
%%%%%%%%%%%%%%%%%%%%%%%%%%%
%\subsection{Relativistic point particle}

%???Maybe we can even tackle this, but even if, I would keep it for a follow-up???

%%%%%%%%%%%%%%%%%%%%%%%%%%%
\subsection{Ehrenfest principle and classical equation of motion}
Let us consider a generic functional of the particle trajectory, $\mathcal{F}[\vec{x}(t)]$. Its expectation value, from the path-integral formulation, is given by the formal expression
\begin{eqnarray}
\langle \mathcal{F} \rangle = \int \mathcal{D}x\,e^{\frac{i}{\hbar}S[\vec{x}(t)]}\mathcal{F}[\vec{x}(t)],
\end{eqnarray}
where we consider the generic non-relativistic action for an external potential $V(\vec{x})$
\begin{eqnarray}
S[\vec{x}(t)] = \int_{t_i}^{t_f}\left\{  
\frac{m}{2}\dot{\vec{x}}^2 - V(\vec{x}(t))
\right\}dt.
\label{eq_action}
\end{eqnarray}
If we now consider a small variation of the tajectory $\vec{x}(t) \rightarrow \vec{x}(t) + \vec{\eta}(t)$, the measure of the path integral $\mathcal{D}x$ remains invariant, and hence we have (expanding up to first order in $\vec{\eta}(t)$)
\begin{eqnarray}
\langle \mathcal{F} \rangle &=& \int \mathcal{D}x\,e^{\frac{i}{\hbar}S[\vec{x}(t)+\vec{\eta}(t)]}\mathcal{F}[\vec{x}(t)+\vec{\eta}(t)]\nonumber\\
&=& \int \mathcal{D}x\,e^{\frac{i}{\hbar}S[\vec{x}(t)]}\left\{
\mathcal{F}[\vec{x}(t)] + \int ds\vec{\eta}(s)\cdot\frac{\delta\mathcal{F}}{\delta\vec{x}(s)} + \frac{i}{\hbar}\mathcal{F}[\vec{x}(t)]\int ds\vec{\eta}(s)\cdot\frac{\delta S}{\delta\vec{x}(s)}
\right\}
\end{eqnarray}
Upon separating the three terms on the right hand side, and exchanging the order of integration, we recover expression
\begin{eqnarray}
\langle \mathcal{F} \rangle = 
\langle \mathcal{F} \rangle + \int ds \vec{\eta}(s)\cdot \left[ \langle \frac{\delta\mathcal{F}}{\delta\vec{x}(s)} \rangle + \frac{i}{\hbar} \langle \frac{\delta S}{\delta\vec{x}(s)} \mathcal{F}\rangle \right]
\end{eqnarray}
Cancelling out the term $\langle \mathcal{F} \rangle$ on both sides, and considering that the variation $\vec{\eta}(s)$ is arbitrary, we conclude
\begin{eqnarray}
\langle \frac{\delta\mathcal{F}}{\delta\vec{x}(s)} \rangle = -\frac{i}{\hbar} \langle \frac{\delta S}{\delta\vec{x}(s)} \mathcal{F}\rangle
\label{eq_F}
\end{eqnarray}
For the action defined in Eq.\eqref{eq_action}, the variation is (for fixed initial $\vec{x}(t_i) =\vec{x}_i$ and final $\vec{x}(t_f) = \vec{x}_f$ points)
\begin{eqnarray}
\frac{\delta S}{\delta\vec{x}(s)} = - m\ddot{\vec{x}}(s) - \nabla V(\vec{x}(s))
\end{eqnarray}
which substituted into Eq.\eqref{eq_F} yields
\begin{eqnarray}
\langle \frac{\delta\mathcal{F}}{\delta\vec{x}(s)} \rangle = \frac{i}{\hbar} \langle \left\{m\ddot{\vec{x}}(s) + \nabla V(\vec{x}(s)) \right\} \mathcal{F}\rangle
\label{eq_FS}
\end{eqnarray}
In particular, for the choice $\mathcal{F} = 1$ into Eq.\eqref{eq_FS}, we recover the classical limit of the equation of motion at the level of the expectation values, a manifestation of Ehrenfest's principle
\begin{eqnarray}
m \langle \ddot{\vec{x}}(s)\rangle = -\langle\nabla V(\vec{x}(s))\rangle
\end{eqnarray}
Even though the previous argument was formulated via the measure $\mathcal{D}\vec{x}$ of the path integral in coordinates space, the same conclusion follows if, as discussed in the previous section, we change the integration variables to incorporate the random velocity kicks between consecutive time steps, i.e. $\vec{x}_{k+1} = \vec{x}_k + \vec{\gamma}_k\delta t$, and $d^3\vec{x}_k = (\delta t)^3 d^3\vec{\gamma}_k$, which in the continuum limit is performed via the jacobian functional determinant that rescales the measure $\mathcal{D}x\rightarrow \mathcal{D}\Gamma$. 

Let us now consider the relativistic covariant action for the single-particle
\begin{eqnarray}
S = -m c \int_{i}^{f}\sqrt{g_{\mu\nu}\frac{dx^{\mu}}{d\tau}\frac{dx^{\nu}}{d\tau}}d\tau.
\label{eq_Srel}
\end{eqnarray}
Here, closely following the argument in Section C, we shall assume that the metric is a superposition of a deterministic classical background and a random local fluctuation, i.e. $g^{\alpha\beta}(x) = g_{BG}^{\alpha\beta}(x) + \delta g^{\alpha\beta}(x)$. For this purpose, it is more convenient to write Eq.\eqref{eq_FS} directly as a general relation between functional variations
\begin{eqnarray}
\langle \delta\mathcal{F}\rangle = -\frac{i}{\hbar}\langle \delta S\,\mathcal{F}\rangle
\label{eq_FS2}
\end{eqnarray}
Now, following the usual procedure to variate the relativistic action in Eq.~\eqref{eq_Srel} that involves the variation of the background metric $g_{BG}^{\mu\nu}(x)$ due to the variation of the trajectories $\delta x^{\mu}$ (with fixed endpoints, see Appendix for algebraic details), one obtains
\begin{eqnarray}
\delta S &=&  mc\int_{i}^{f}\left\{
g_{BG,\mu\nu}\frac{d^2 x^{\nu}}{d\tau^2} + \frac{1}{2}\frac{dx^{\alpha}}{d\tau}\frac{dx^{\nu}}{d\tau}\left( \partial_{\alpha}g_{BG,\mu\nu} +
\partial_{\nu}g_{BG,\mu\alpha} - \partial_{\mu}g_{BG,\alpha\nu}
\right)
\right\}\delta x^{\mu}d\tau\nonumber\\
&=&  mc\int_{i}^{f}\left\{
g_{BG,\mu\beta}\left(\frac{d^2 x^{\beta}}{d\tau^2} + \Gamma_{\,\alpha\nu}^{\beta}\frac{dx^{\alpha}}{d\tau}\frac{dx^{\nu}}{d\tau}\right)
\right\}\delta x^{\mu}d\tau,
\label{eq_Svar}
\end{eqnarray}
where in the last line we identified the Christoffel symbol for the background metric $g^{\mu\nu}_{BG}$
\begin{eqnarray}
\Gamma_{\,\alpha\nu}^{\beta} = \frac{1}{2}g_{BG}^{\mu\beta}\left(
\partial_{\alpha}g_{BG,\mu\nu} + \partial_{\nu}g_{BG,\mu\alpha} - \partial_{\mu}g_{BG,\alpha\nu}
\right).
\end{eqnarray}
By inserting Eq.~\eqref{eq_Svar} into Eq.~\eqref{eq_FS2}, and choosing as before $\mathcal{F} = 1$, we obtain
\begin{eqnarray}
\langle \delta S \rangle = 0
=  m c \int_{i}^{f} \left\langle
g_{BG,\mu\beta}\left(\frac{d^2 x^{\beta}}{d\tau^2} + \Gamma_{\,\alpha\nu}^{\beta}\frac{dx^{\alpha}}{d\tau}\frac{dx^{\nu}}{d\tau}\right)\right\rangle\delta x^{\mu}d\tau,
\end{eqnarray}
which implies the geodesic equation at the level of the expectation value
\begin{eqnarray}
\left\langle
g_{BG,\mu\beta}\left(\frac{d^2 x^{\beta}}{d\tau^2} + \Gamma_{\,\alpha\nu}^{\beta}\frac{dx^{\alpha}}{d\tau}\frac{dx^{\nu}}{d\tau}\right)\right\rangle = 0,
\end{eqnarray}
in agreement with Ehrenfest's principle, just as in the non-relativistic case.
%%%%%%%%%%%%%%%%%%%%%%%%%%%%
\subsection{Discussion}

When realizing our proofs, we basically just performed a series of changes of variables, re-definitions, and mathematical identities to the usual PI quantization. Further, 
the integrals over connections always involve the previous and subsequent position and thus, a multi-step application 
can make the algebra more cumbersome.
\begin{center}
    ``{\it {So what is the point''?}}
\end{center}
The benefit and new insight is 
not in the mathematical steps, but in the largely different conceptual meaning:
The EQQ in flat space-time 
can be interpreted as an integral over geodesics
in a space-time with virtual deformations at intermediate times.
These deformations occur at all points in space-time, but according to the definition (\ref{eq_DPhi}) they are reduced to a functional integral over the fluctuating component of the connections $\delta\Gamma$, that hence may be interpreted in the low-curvature Newtonian limit as an effective stochastic force.
Thus, EQQ shares this idea with the
stochastic quantization method, 
where also vacuum or spacetime fluctuations are the reason for quantum mechanics~\cite{Nelson:1966sp,Kuipers:2023pzm,Kuipers:2023ibv},
even though the realization is different.

As they stand, all of these results are just a change of perspective of things that are already known.
Below, we want comment on possible consequences of this ``change of perspective'' for other approaches and future developments. Naturally, those comments about the future are more far fetched. We recommend the reader to take them with a good measure of scepticism and care:

Relating two different functional integrals in this way, e.g. with a functional delta like in (\ref{eq_PIEQ})
will have consequences for all programs which attempt to quantize geometric degrees of freedom $\Gamma$ and matter fields $\phi$~\cite{tHooft:1974toh,Dona:2013qba}, e.g. in terms
of seemingly independent functional integrals 
\be\label{eq_QG2}
\int {\mathcal{D}}\Gamma \int {\mathcal{D}}\phi \;e^{\frac{i}{\hbar}S}\dots 
\ee
Imposing a $\delta[\cdot]$ restriction in (\ref{eq_QG2}) will reduce the functional measure and eventually help to avoid infinities. This is reminiscent of the mechanism that renders delta gravity finite~\cite{Alfaro:2010it,Alfaro:2013ega}.
Further similarities of our proposal exist to a line of research which 
explores the possibility of understanding quantum mechanics in terms of geometric concepts~\cite{Carinena:2007,Koch:2010bz}.

In addition to this, there are other intriguing and philosophical aspects of the EQQ that we'd like to touch on.
\begin{itemize}
    \item Background resolution vs. particle resolution:\\
    The first comment comes from the hypothesis that the motion of a quantum particle is caused by fluctuating local curvatures. 
This motion only appears to be 
erratic, since it is perceived from the macroscopic perspective of a flat background.
The perspective of a flat background is accounted for by the fact that the actions, which enter in the exponential weight of the curvature integrals~(\ref{eq_PIEQ4}), are actually free actions in flat space-time. If it would be possible to increase the $\delta t$ resolution below the scale of the space-time fluctuations (the Planck scale?), then the measurement would be aware of the local changes $\vec \gamma_i$. This would imply that
the local experiment
 would actually co-move with 
 the point particle. In this extreme case, the exponential action should also contain the effects of the local $\vec \Gamma_i$.
 \item
 Equivalence principle (EP):\\
 The second comment relates to the EP, which states that ``{\it{the effects of gravity in terms of a particular background metric and the effects of acceleration are indistinguishable}}''~\cite{Misner:1973prb}. 
 By many, the EP is seen
 as the conceptual corner stone of GR.
 Vexatiously, the EP in this form is already in conflict with simple quantum mechanics (QM), even at ``mesoscopic'' distance scales (scales far bigger than the Planck length, but small enough to be sensitive to quantum effects).
For example, one of the challenges in reconciling the EP with QM is the fact that quantum mechanics allows for the possibility of superpositions, which are combinations of different states that can interfere with each other. This can lead to scenarios where the 
two particle states, such as being at rest or being accelerated are superimposed at mesoscopic scales, while the above EP needs to relate to a metric and thus
demands to pick one of the two states. 
The obvious way out of this dilemma would be to allow for metric superpositions at mesoscopic scales. This, however, could mean to release the beast of quantum gravity at distance scales much larger than the Planck scale.
There are numerous attempts to reformulate the EP, to a quantum version (QEP)
in such a way that it 
is at least compatible with QM at mesoscopic scales 
\cite{Lammerzahl:1998pt,Giacomini:2020ahk,Giacomini:2021aof}. 
The EQQ approach might add to this discussion since it can be cast into 
the statement: ``{\it{The random motion of PI's is indistinguishable from geodesic motion caused by random gravitational fluctuations}}''.
This is a new candidate for a QEP since
the classical limit of QM paths leads to 
classical paths with accelerations and the classical limit of gravitational fluctuations leads to macroscopic classical curvature.
Thus, we are tempted to conjecture
that the macroscopic classical limit of this QEP-candidate is the usual EP.
Even more, this QEP-candidate has the advantage, that it is formulated at the level of paths in the amplitude. As such, it naturally allows for superposition and avoids the usual conflicts of the classical EP with 
superposition and non-locality of the wave function. Interestingly, it has recently been argued that the evolution of states in a superposition of spacetimes is
equivalent to an evolution of superposed states in classical spacetime~\cite{Foo:2023vbr}.
\end{itemize}

%%%%%%%%%%%%%%%%%%%%%%%%%%%
\section{Conclusion and outlook}

We have shown, that it is possible to 
formulate non-relativistic QM in terms of
an integral over geodesic paths on a random background instead of
an integral over random free paths on a flat background.
This novel perspective introduces a wealth of new questions and opportunities for further exploration, including the generalization to systems with many particles, particles possessing spin, relativistic point particles~\cite{Koch:2017nha,Koch:2017bvv,Koch:2019vxw,Koch:2020dql}, quantum field theory, and even the coupling of quantum gravity with matter.
%%%%%%%%%%%%%%%%%%%%%%%%%
\section{Acknowledgements}

E.M. acknowledges financial support from the project ANID PIA Anillo ACT/192023, and project Fondecyt 1230440.
We thank D. Grumiller for feedback on the draft.
%%%%%%%%%%%%%%%%%%%%%%%%%
\appendix
\section{Functional determinants}
In this section, we shall present a detailed calculation of two functional determinants that emerge in the explicit examples presented in the text. Such calculation involves the solution of the Sturm-Liouiville eigenvalue problem associated to each linear differential operator.
\subsection{Free particle case}
We seek for the eigenvalues $\lambda$ and eigenfunctions $\psi_{\lambda}(t)$ of the Sturm-Liouiville problem
\bea
-m\frac{d^2}{dt^2}\psi_{\lambda}(t) &=& \lambda \psi_{\lambda}(t)\nonumber\\
\psi_{\lambda}(t_i)&=& 
\psi_{\lambda}(t_f)= 0.
\eea
The general solution can be written as the linear combination
\be
\psi_{\lambda}(t) = A_{\lambda}\sin\left(  \sqrt{\lambda/m}(t-t_i)\right) + B_{\lambda}\cos\left(  \sqrt{\lambda/m}(t-t_i)\right).
\ee
Now, applying the first boundary condition leads to
\be
\psi_{\lambda}(t_i) = B_{\lambda} = 0,
\ee
while subsequently applying the second boundary condition
\be
\psi_{\lambda}(t_f) = A_{\lambda}\sin\left(  \sqrt{\lambda/m}(t_f-t_i)\right) = 0.
\ee
This equation possesses nontrivial solutions $A_{\lambda}\ne 0$ if
\be
\sqrt{\lambda_n/m} = \frac{n\pi}{t_f - t_i},
\ee
for $n = 1, 2, \ldots,\infty$.
Therefore, the formal expression for the associated functional determinant is given by the product of such eigenvalues, to the power of the dimensionality of the identity ($d=3$ in our examples)
\be
{\mathrm{Det}}\left[-\mathbf{1}m\frac{d^2}{dt^2} \right] = \left[ \prod_{n=1}^{\infty} \lambda_{n}\right]^3 = \left[ \prod_{n=1}^{\infty} m\left(\frac{n\pi}{t_f-t_i} \right)^2\right]^3.
\label{eq_det_fp}
\ee
This expression is clearly divergent, and hence the importance of the role of the overall normalization factor in the path-integral measure. Nevertheless, as we shall see in the second example, a ratio between two such functional determinants indeed leads to a well defined, finite limit.
\subsection{The harmonic potential}
In this second example, we seek for the eigenvalues $\lambda$, and corresponding eigenfucntions $\psi_{\lambda}(t)$, for the Sturm-Liouville problem
\bea
-m\frac{d^2}{dt^2}\psi_{\lambda}(t) - m\Omega^2\psi_{\lambda}(t) &=& \lambda \psi_{\lambda}(t)\nonumber\\
\psi_{\lambda}(t_i)&=& 
\psi_{\lambda}(t_f)= 0.
\eea
The general solution is now given by the linear combination
\be
\psi_{\lambda}(t) = A_{\lambda}\sin\left(  \sqrt{\Omega^2+\lambda/m}(t-t_i)\right) + B_{\lambda}\cos\left(  \sqrt{\Omega^2+\lambda/m}(t-t_i)\right).
\ee
Now, applying the first boundary condition, we obtain
\be
\psi_{\lambda}(t_i) = B_{\lambda} = 0.
\ee
Now, by applying the second boundary condition, we are lead to the equation
\be
\psi_{\lambda}(t_f) = A_{\lambda}\sin\left(  \sqrt{\Omega^2+\lambda/m}(t_f-t_i)\right) = 0.
\ee
This equation possesses nontrivial solutions $A_{\lambda}\ne 0$ if
\be
\sqrt{\Omega^2+\lambda_n/m} = \frac{n\pi}{t_f - t_i},
\ee
for $n = 1, 2, \ldots,\infty$.
Therefore, the formal expression for the associated functional determinant is given in this case by
\be
{\mathrm{Det}}\left[\mathbf{1}\left(-m\frac{d^2}{dt^2} -m\Omega^2\right)\right] = \left[ \prod_{n=1}^{\infty} \lambda_{n}\right]^3 = \left[ \prod_{n=1}^{\infty} \left\{m\left(\frac{n\pi}{t_f-t_i} \right)^2-m\Omega^2\right\}\right]^3.
\label{eq_det_fp}
\ee
We notice that this is again a divergent quantity. Nevertheless, after the overall normalization of the path-integral is defined by Eq.~\eqref{eq_Knorm}, we just need the ratio between the two functional determinants, which is
\bea
\frac{\left({\rm{Det}}\left[ \mathbf{1}\left(-m\frac{d^2}{dt^2} - m\Omega^2\right)  \right]\right)^{-1/2}}{\left({\rm{Det}}\left[ \mathbf{1}\left(-m\frac{d^2}{dt^2} \right)  \right]\right)^{-1/2}} &=&
\left[\prod_{n=1}^{\infty}\left\{
\frac{m\left(\frac{n\pi}{t_f-t_i} \right)^2-m\Omega^2}{m\left(\frac{n\pi}{t_f-t_i} \right)^2}
\right\}\right]^{-3/2}
= \left[\prod_{n=1}^{\infty}\left\{1 - 
\frac{\left(\Omega (t_f - t_i)/\pi\right)^2}{n^2}
\right\}\right]^{-3/2}.
\eea
Clearly, this ratio is now a convergent product, after the general identity
\be
\frac{\sin(\pi z)}{\pi z} = \prod_{n=1}^{\infty}\left\{ 
1 - \frac{z^2}{n^2}
\right\}.
\ee
Therefore, we finally obtain the desired result
\be
\frac{\left({\rm{Det}}\left[ \mathbf{1}\left(-m\frac{d^2}{dt^2} - m\Omega^2\right)  \right]\right)^{-1/2}}{\left({\rm{Det}}\left[ \mathbf{1}\left(-m\frac{d^2}{dt^2} \right)  \right]\right)^{-1/2}}
=\left( \frac{\sin\left( \Omega(t_f - t_i)\right)}{\Omega (t_f - t_i)} \right)^{-3/2}.
\ee

\section{Variation of the relativistic action for the point particle}
The action of the relativistic point particle is invariant under under parameterization of the trajectories, so we can write (for the backgorund metric $g_{BG,\mu\nu}(x)$)
\begin{eqnarray}
S = - m c\int_{i}^{f}\sqrt{g_{BG,\mu\nu}(x)\frac{dx^{\mu}}{d\lambda}\frac{dx^{\nu}}{d\lambda}}d\lambda
\label{eq_A1}
\end{eqnarray}
Performing a variation, we have
\begin{eqnarray}
\delta S = -\frac{mc}{2}\int_{i}^{f}\frac{\delta\left( g_{BG,\mu\nu}(x)\frac{dx^{\mu}}{d\lambda}\frac{dx^{\nu}}{d\lambda} \right)}{\sqrt{g_{BG,\mu\nu}(x)\frac{dx^{\mu}}{d\lambda}\frac{dx^{\nu}}{d\lambda}}}d\lambda
\label{eq_A2}
\end{eqnarray}
By defining the proper time $d\tau$, we have
\begin{eqnarray}
\frac{d\lambda}{d\tau} = \frac{1}{\sqrt{g_{BG,\mu\nu}(x)\frac{dx^{\mu}}{d\lambda}\frac{dx^{\nu}}{d\lambda}}},
\label{eq_A3}
\end{eqnarray}
such that Eq.~\eqref{eq_A2} becomes
\begin{eqnarray}
\delta S &=& - \frac{mc}{2}\int_{i}^{f}\delta\left( g_{BG,\mu\nu}(x)\frac{dx^{\mu}}{d\lambda}\frac{dx^{\nu}}{d\lambda} \right)\frac{d\lambda}{d\tau}\,d\lambda\nonumber\\
&=& - \frac{mc}{2}\int_{i}^{f} \left[
\frac{dx^{\mu}}{d\lambda}\frac{dx^{\nu}}{d\tau}\delta g_{BG,\mu\nu} +
2 g_{BG,\mu\nu}\frac{d\left( \delta x^{\mu} \right)}{d\lambda}\frac{dx^{\nu}}{d\tau}
\right]d\lambda
\label{eq_A4}
\end{eqnarray}
The variation of the background metric is given by $\delta g_{BG,\mu\nu}(x) = \partial_{\alpha}g_{BG,\mu\nu}\,\delta x^{\alpha}$, such that we can write Eq.~\eqref{eq_A4} as (upon changing the integration variable $\lambda\rightarrow\tau$)
\begin{eqnarray}
\delta S &=& - \frac{mc}{2}\int_{i}^{f} \left[
\frac{dx^{\mu}}{d\tau}\frac{dx^{\nu}}{d\tau}\,\partial_{\alpha} g_{BG,\mu\nu}\,\delta x^{\alpha}
- 2\delta x^{\mu}\frac{d}{d\tau}
\left(g_{BG,\mu\nu} \frac{dx^{\nu}}{d\tau}  \right)
+ \frac{d}{d\tau}\left( 2 g_{BG,\mu\nu}\,\delta x^{\mu}\frac{dx^{\nu}}{d\tau}  \right)
\right]d\tau.
\label{eq_A5}
\end{eqnarray}
The last term on the right hand side of Eq.~\eqref{eq_A5} is a boundary term, that vanishes if the endpoints of the trajectory remain fixed, i.e. $\delta x^{\mu}(i) = 0$, $\delta x^{\mu}(f) = 0$.
Further expanding the derivative in the second term, and collecting common factors, we arrive at 
\begin{eqnarray}
\delta S &=&  mc\int_{i}^{f} \left[
g_{BG,\mu\nu}\frac{d^2 x^{\nu}}{d\tau^2} + \partial_{\alpha}g_{BG,\mu\nu}
\frac{dx^{\alpha}}{d\tau}\frac{dx^{\nu}}{d\tau}
-\frac{1}{2}\partial_{\mu}g_{BG,\alpha\nu}\frac{dx^{\alpha}}{d\tau}\frac{dx^{\nu}}{d\tau}
\right]\delta x^{\mu}d\tau.
\label{eq_A6}
\end{eqnarray}
Now, using the trivial identity
\begin{eqnarray}
\partial_{\alpha}g_{BG,\mu\nu}
\frac{dx^{\alpha}}{d\tau}\frac{dx^{\nu}}{d\tau}
= \frac{1}{2}\left( \partial_{\alpha}g_{BG,\mu\nu}
+ \partial_{\nu}g_{BG,\mu\alpha}
\right)\frac{dx^{\alpha}}{d\tau}
\frac{dx^{\nu}}{d\tau},
\label{eq_A7}
\end{eqnarray}
into Eq.~\eqref{eq_A6}, we finally obtain the result presented in Eq.~\eqref{eq_Svar} in the main text
\begin{eqnarray}
\delta S = m c \int_{i}^{f}\left[
g_{BG,\mu\nu}\frac{d^{2}x^{\nu}}{d\tau^2} + \frac{1}{2}\frac{dx^{\alpha}}{d\tau}
\frac{dx^{\nu}}{d\tau}\left(
\partial_{\alpha}g_{BG,\mu\nu}
+ \partial_{\nu}g_{BG,\mu\alpha}
-\partial_{\mu}g_{BG,\alpha\nu}
\right)
\right]\delta x^{\mu}d\tau
\end{eqnarray}
%%%%%%%%%%%%%%%%%%%%%%%%%%%%%%%%


\begin{thebibliography}{xxxx}
\bibitem{EQQ:2023}
Note that this name is not meant as trivialization of the devastating effects
an earthquake can provoke in inhabited regions.
Instead, we encourage everyone to donate to those who are helping in the aftermath of such an event.
\end{thebibliography}
\end{document}